\documentclass[11pt,a4paper]{article}

\usepackage[margin=1in]{geometry}
\usepackage{setspace}
\usepackage{amsmath,amssymb,amsthm}
\usepackage{mathtools}
\usepackage{enumerate}
\usepackage{xcolor}
\usepackage[noblocks]{authblk}
\usepackage[round,authoryear]{natbib}
\usepackage{hyperref}
\hypersetup{colorlinks=true,linkcolor=blue,citecolor=blue,urlcolor=blue}
\usepackage[capitalise,noabbrev,nameinlink]{cleveref}

\newtheorem{theorem}{Theorem}[section]
\newtheorem{proposition}[theorem]{Proposition}
\newtheorem{lemma}[theorem]{Lemma}
\newtheorem{corollary}[theorem]{Corollary}
\theoremstyle{definition}
\newtheorem{definition}[theorem]{Definition}
\newtheorem{remark}[theorem]{Remark}
\newtheorem{example}[theorem]{Example}

\newcommand{\R}{\mathbb{R}}

\newcommand{\E}{\mathbb{E}}
\newcommand{\e}{\mathrm{e}}
\newcommand{\A}{\mathcal{A}}
\newcommand{\HH}{\mathcal{H}}
\newcommand{\X}{\mathcal{X}}
\newcommand{\bx}{\mathbf{x}}
\newcommand{\bP}{\mathbf{P}}
\newcommand{\bzero}{\mathbf{0}}
\DeclareMathOperator{\Span}{span}
\DeclareMathOperator{\ran}{ran}
\DeclareMathOperator{\diag}{diag}
\DeclareMathOperator*{\argmin}{arg\,min}
\numberwithin{equation}{section}

\title{Drift-invariant kernels for discount curve models}

\author[3]{Andreas Celary}
\author[1]{Zehra Eksi Altay\textsuperscript{*}}
\author[2]{Damir Filipovic}
\author[1]{Paul Kr\"uhner}
\affil[1]{Vienna University of Economics and Business, Welthandelsplatz 1, 1020 Vienna, Austria. Emails: \texttt{zehra.eksi@wu.ac.at}, \texttt{paul.eisenberg@wu.ac.at}}
\affil[2]{EPFL, 1015 Lausanne, Switzerland. Email: \texttt{damir.filipovic@epfl.ch}}
\affil[3]{Princeton University, Princeton, New Jersey 08544, USA. Email: \texttt{acelary@princeton.edu}}

\date{November 5, 2026}  %% adjust at each revision step

\begin{document}

\maketitle
\begingroup
\renewcommand{\thefootnote}{*}
\footnotetext{Corresponding author: Zehra Eksi Altay.
Email: \texttt{zehra.eksi@wu.ac.at}.}
\endgroup

\begin{abstract}
Kernel ridge regression estimates the discount curve as the minimizer of a pricing error plus a penalty given by the norm of a reproducing kernel Hilbert space. By the representer theorem, the estimate lies in a finite-dimensional affine space that depends only on the kernel and on the cash flow dates of the instruments. We ask when this space is invariant under the risk-neutral drift of the discount curve, so that the estimated curves can be the states of a continuous-time arbitrage-free model. We show that this holds for every grid of cash flow dates if and only if the kernel is of the form $k(x,y)=\e^{b(x+y)}\big(c\,\e^{axy}+1-c\big)$, apart from kernels whose Hilbert space has dimension at most two. For this kernel the estimated curve is an affine combination of flat discount curves, each corresponding to a constant continuously compounded yield. The drift-invariant subspaces correspond to sub-grids of the cash flow dates, and the discounted coefficients have linear risk-neutral dynamics.
\end{abstract}



\section{Introduction}

The discount curve at time $t$ is the function $x\mapsto p_t(x)=P(t,t+x)$, where $P(t,T)$ denotes the price at time $t$ of a zero-coupon bond with maturity $T$. The discount curve is not observed at all maturities. It must be estimated from the prices of traded instruments with cash flows at finitely many dates. This estimation is typically performed separately on each observation date, but the resulting curves may also serve as input to dynamic term structure models.

Kernel ridge regression (KRR) estimates the curve as the minimizer of a weighted pricing error plus a penalty given by the norm of a reproducing kernel Hilbert space (RKHS), see \cite{filipovic3,filipovic_kr}. By the representer theorem, the estimate lies in the finite-dimensional affine space
\begin{equation}\label{eq:A_intro}
  \A = p_0 + \Span\{k_0(x_1,\cdot),\dots,k_0(x_N,\cdot)\},
\end{equation}
where $x_1,\dots,x_N$ are the cash flow dates, $p_0$ is a prior curve, and $k_0$ is the Schur complement of the kernel at the point $0$, which is the reproducing kernel of the functions vanishing at $0$. The space $\A$ depends on the kernel and on the cash flow dates only, not on the prices.

A good fit to observed prices on individual dates does not, by itself, guarantee that the resulting curves have arbitrage-free dynamics. A continuous-time dynamic model for the discount curve has to satisfy the no-arbitrage drift condition of \cite{filipovic_discount}: under a risk-neutral measure the drift of $p_t$ equals $\eta(p_t)=p_t'-p_t'(0)\,p_t$. If the curve is re-estimated every day with the same kernel on a fixed grid of cash flow dates, all estimates lie in $\A$. A dynamic model that reproduces them must keep the curve in $\A$. This requires that $\eta$ maps $\A$ into its direction space. We call such $\A$ drift-invariant. This is the consistency problem of \cite{bjoerk1,filipovic_consistency}, posed for discount curves instead of forward curves, and for the spaces produced by KRR instead of parametric families.

Our main result, \cref{thm:main}, classifies the kernels for which $\A$ is drift-invariant for every grid of cash flow dates. Apart from kernels whose RKHS has dimension at most two, these are the kernels
\begin{equation}\label{eq:kernel_intro}
  k(x,y) = \e^{b(x+y)}\big(c\,\e^{axy}+1-c\big)\quad\text{with $a,c>0$ and $b\in\R$}.
\end{equation}
The kernel $\e^{axy}$ is the Segal--Bargmann kernel \cite{Bargmann62}. Its RKHS consists of real analytic functions, and its norm penalizes Taylor coefficients rather than derivatives. The parameter $c$ is a scale: it does not affect $\A$, and it enters the estimator only as the prior variance of the curve in the Gaussian process view of KRR. Drift-invariance is a condition on continuous-time dynamics, and it is of a different nature than the smoothness rewards used for static curve estimation in \cite{filipovic3,filipovic_kr}. See the discussion after \cref{thm:main}.

For the kernel \eqref{eq:kernel_intro} the structure of $\A$ is explicit, with several consequences for term structure modeling. First, $\A$ consists of the affine combinations of the flat discount curves $\e^{(ax_j+b)x}$, $j=0,\dots,N$, with $x_0=0$, each corresponding to a constant continuously compounded yield (\cref{lem:expspan}). The kernel-based estimator thus generates a finite-dimensional family with an explicit financial interpretation. Second, the drift-invariant affine subspaces of $\A$ are those spanned by sub-grids of the cash flow dates (\cref{prop:subgrid}). In particular, dimension reductions based on principal components of the kernel matrix do not preserve drift-invariance (\cref{cor:pca}). Third, the coefficient process of a discount curve model in $\A$ has a quadratic drift under the risk-neutral measure, and the discounted coefficients have a linear drift (\cref{thm:dyn}). This places the model in the class of linear-rational term structure models of \cite{filipovic_lrtsm} (\cref{rem:lr}), and it yields well-defined multi-factor models by specifying the volatility of the discounted coefficients.

\emph{Related literature.} Static curve fitting uses parametric families \cite{nelson1987parsimonious,svensson1994estimating} or splines \cite{mcculloch1971measuring,vasicek1982exponential,adams1994fitting,fisher1995fitting}. See \cite{haganwest2006interpolation} for an overview. The consistency of parametric families with arbitrage-free forward curve dynamics was studied in \cite{bjoerk1,filipovic1999note,filipovic_exponential,filipovic_consistency}, and finite-dimensional realizations in \cite{bjoerk_svensson,bjoerk,teichmann1,teich,tappe4,tappe2}. The drift condition for discount curves is derived in \cite{filipovic_discount}. Dynamically consistent forward curve fitting is proposed in \cite{jarrow_wu}, consistent recalibration in \cite{richter_teichmann}, and data-driven curve models in \cite{autoencoder}. The paper \cite{shijie} shows that statistical consistency of term structure estimates forces an affine geometry. An empirical implementation of a related RKHS approach, using US Treasury data, is developed in \cite{celary_rkhs}. The exponential kernel employed there is a special case of the family \eqref{eq:kernel_intro}. While \cite{celary_rkhs} focuses on curve estimation, dimension reduction and model calibration, the present paper characterizes the kernels whose hypothesis spaces are drift-invariant for every grid of cash flow dates. We do not study empirical performance here.

\emph{Outline.} \cref{sec:drift} recalls the drift condition and characterizes drift-invariant affine spaces. \cref{sec:krr} contains the main result and the structure of $\A$. \cref{sec:dyn} characterizes the coefficient dynamics under the no-arbitrage drift condition. \cref{sec:conclusion} concludes. All proofs are in the appendix: \cref{app:interp} characterizes the interpolating subspaces of an RKHS, which is used in \cref{sec:krr}, and \cref{app:proofs} contains the proofs of the results of \cref{sec:drift,sec:krr,sec:dyn}, including the classification of drift-invariant kernels on which \cref{thm:main} rests.

\emph{Notation.} $\R_+=[0,\infty)$. $C^1(\R_+)$ is the space of continuously differentiable functions $f:\R_+\to\R$, with one-sided derivative $f'(0)$ at $0$. A kernel is a symmetric positive semidefinite function $k:\R_+\times\R_+\to\R$. Its RKHS is denoted by $\HH$, see \cite{aronszajn,paulsen_rkhs}. For $h_1,h_2:\R_+\to\R$ we write $h_1\otimes h_2$ for the function $(x,y)\mapsto h_1(x)h_2(y)$.

\section{Drift-invariant affine spaces}\label{sec:drift}

We use the setting of \cite{filipovic_discount}. Let $p_t(x)=P(t,t+x)$ denote the discount curve at time $t$. Under a risk-neutral measure, absence of arbitrage requires that the drift of $P(t,T)$ equals $r_tP(t,T)$, where $r_t=-\partial_xp_t(0)$ is the short rate, see \cite[Section 2]{filipovic_discount}. In terms of the curve $p_t$ this means that $p_t$ has drift $\eta(p_t)$, where
\begin{equation}\label{eq:eta}
  \eta(f) = f' - f'(0)\,f,\qquad f\in C^1(\R_+),
\end{equation}
see \cref{sec:dyn} below for the precise statement.

\begin{definition}\label{def:inv}
Let $A=p_0+V\subseteq C^1(\R_+)$ be an affine space with base point $p_0$ and direction space $V$. We call $A$ \emph{drift-invariant} if $f(0)=1$ and $\eta(f)\in V$ for all $f\in A$.
\end{definition}

\begin{proposition}\label{prop:inv}
Let $A=p_0+V\subseteq C^1(\R_+)$ be an affine space and define the linear map
\begin{equation}\label{eq:alpha}
  \alpha(h) = h' - h'(0)\,p_0,\qquad h\in V.
\end{equation}
The following are equivalent:
\begin{enumerate}[(1)]
\item $A$ is drift-invariant;
\item $p_0(0)=1$, $h(0)=0$ for all $h\in V$, $\eta(p_0)\in V$, and $\alpha(V)\subseteq V$.
\end{enumerate}
\end{proposition}

The proof is in \cref{app:proofs}. Condition (2) separates the role of the base point from that of the direction space: $\eta(p_0)\in V$ is a condition on $p_0$, and $\alpha(V)\subseteq V$ is a linear condition on $V$.

\section{Kernel ridge regression and drift-invariance}\label{sec:krr}

In this section we describe the affine space in which the KRR estimate of the discount curve lies, and we characterize the kernels for which this space is drift-invariant. Throughout this section, $k$ is a kernel on $\R_+$ that is continuous and continuously differentiable in each variable. We also assume $k(0,0)=1$. This is a normalization: if $k(0,0)>0$ we can rescale $k$, which changes the estimator below only through its regularization parameter, and if $k(0,0)=0$ then $h(0)=0$ for all $h\in\HH$ and $\HH$ contains no discount curve.

\subsection{The hypothesis space}

Discount curves satisfy $p(0)=1$. The following lemma describes the discount curves in $\HH$. It is the case $m=1$, $x_1=0$, $P_1=1$ of \cref{thm:interp} in \cref{app:interp}.

\begin{lemma}\label{lem:H0}
Let $p_0=k(0,\cdot)$ and
\begin{equation}\label{eq:k0}
  k_0(x,y) = k(x,y) - k(x,0)\,k(0,y),\qquad x,y\in\R_+ .
\end{equation}
Then $\HH_0=\{h\in\HH: h(0)=0\}$ is a closed subspace of $\HH$, it is the orthogonal complement of $\Span\{p_0\}$, and it is an RKHS with kernel $k_0$. Moreover,
\begin{equation}\label{eq:affine}
  \{h\in\HH: h(0)=1\} = p_0 + \HH_0,\qquad \|p_0+h_0\|_\HH^2 = 1+\|h_0\|_{\HH_0}^2\quad\text{for } h_0\in\HH_0 .
\end{equation}
\end{lemma}

The kernel $k_0$ is the Schur complement of $k$ at the point $0$. In Gaussian process terms, it is the conditional covariance given the value at $0$.

We now recall KRR for the discount curve, see \cite{filipovic3,filipovic_kr}. On the estimation date we observe the prices $P_1,\dots,P_m$ of $m$ instruments. Instrument $i$ pays the cash flows $c_{i1},\dots,c_{iN}$ at the dates $0<x_1<\dots<x_N$, measured in time to maturity. Let $w_1,\dots,w_m>0$ be weights and $\lambda>0$ a regularization parameter. The KRR estimate of the discount curve is
$\hat p=p_0+\hat h$ with
\begin{equation}\label{eq:krr}
  \hat h = \argmin_{h\in\HH_0}\Big\{ \sum_{i=1}^m w_i\Big(P_i-\sum_{j=1}^N c_{ij}\,(p_0+h)(x_j)\Big)^2 + \lambda\,\|h\|_{\HH_0}^2\Big\}.
\end{equation}
This is the KRR of \cite{filipovic3} with prior curve $p_0$ and kernel $k_0$. By \eqref{eq:affine}, $\hat p$ is also the minimizer of the pricing error plus $\lambda\|g\|_\HH^2$ over all $g\in\HH$ with $g(0)=1$. The solution is explicit, see \cite[Theorem A.1]{filipovic3}: with $\bx=(x_1,\dots,x_N)^\top$, $K_0=(k_0(x_i,x_j))_{i,j=1}^N$, $C=(c_{ij})\in\R^{m\times N}$, $P=(P_1,\dots,P_m)^\top$ and $\Lambda=\diag(\lambda/w_1,\dots,\lambda/w_m)$,
\begin{equation}\label{eq:beta}
  \hat h=\sum_{j=1}^N\beta_j\,k_0(x_j,\cdot),\qquad \beta = C^\top\big(CK_0C^\top+\Lambda\big)^{-1}\big(P-Cp_0(\bx)\big),
\end{equation}
where $p_0(\bx)=(p_0(x_1),\dots,p_0(x_N))^\top$. Hence $\hat p$ lies in the affine space
\begin{equation}\label{eq:A}
  \A = p_0 + V,\qquad V=\Span\{k_0(x_1,\cdot),\dots,k_0(x_N,\cdot)\},
\end{equation}
for any prices and weights. We call $\A$ the hypothesis space of the KRR for the grid $x_1<\dots<x_N$. Since $k_0(x_j,0)=0$, every $f\in\A$ satisfies $f(0)=1$. Since $k$ is continuously differentiable in each variable, $\A\subseteq C^1(\R_+)$.

In practice the grid consists of the cash flow dates of the instruments on the estimation date. These dates, measured in time to maturity, change from day to day. One can also use a fixed grid that contains all cash flow dates, for example all day counts up to thirty years. Then $\A$ is the same space on every day, and the coefficients of $\hat p$ have many zeros. The dimension of $\A$ is then large, of the order of $10^4$, and \cref{prop:subgrid} below describes how it can be reduced.

\begin{definition}\label{def:kinv}
We call the kernel $k$ \emph{drift-invariant} if the hypothesis space \eqref{eq:A} is drift-invariant for every $N\in\{0,1,2,\dots\}$ and every grid $0<x_1<\dots<x_N$. For $N=0$ the grid is empty and $\A=\{p_0\}$.
\end{definition}

\subsection{Main result}

\begin{theorem}\label{thm:main}
The kernel $k$ is drift-invariant if and only if there is $b\in\R$ such that one of the following holds:
\begin{enumerate}[(1)]
\item $k(x,y)=\e^{b(x+y)}\big(c\,\e^{axy}+1-c\big)$ for some $a,c>0$;
\item $k(x,y)=\e^{b(x+y)}\big(1+\varphi(x)\varphi(y)\big)$, where $\varphi(x)=c(\e^{\tilde ax}-1)$ for some $c\ge0$ and $\tilde a\in\R\setminus\{0\}$, or $\varphi(x)=cx$ for some $c\ge0$.
\end{enumerate}
In case (1), $\dim\HH=\infty$. In case (2), $\dim\HH\le2$. In particular, if $\dim\HH\ge3$, then $k$ is drift-invariant if and only if (1) holds.
\end{theorem}

The proof is in \cref{app:proofs}. Case (2) consists of the kernels that are sums of at most two elementary tensors. The form $\varphi(x)=cx$ is the limit $\tilde a\to0$ of the form $\varphi(x)=c(\e^{\tilde ax}-1)$. These kernels are of little interest for curve estimation, since their hypothesis spaces have dimension at most one. The main case is (1).

\cref{thm:main} shows that drift-invariance is a strong requirement. In both cases, $k(x,\cdot)$ is real analytic on $\R$ for every $x$. In case (1) the RKHS of $k$ consists of functions of the form $\e^{bx}f(x)$ with $f$ the restriction of an entire function, and the norm of $f$ in the RKHS of $\e^{axy}$ is $\|f\|^2=\sum_{n\ge0} n! f_n^2/(a^n)$ for $f(x)=\sum_{n\ge0} f_nx^n$, see \cite{Bargmann62} and \cite[Section 4.4.2]{paulsen_rkhs}. The penalty in \eqref{eq:krr} thus acts on the Taylor coefficients of the curve at $0$, and not on its derivatives as in \cite{filipovic3}. Kernels that reward smoothness of the curve, such as the Sobolev type kernels of \cite{filipovic3,filipovic_kr}, are not real analytic across the diagonal. They are designed for the static estimation problem on a given date, and they are not drift-invariant in the sense of \cref{def:kinv}. Drift-invariance concerns the continuous-time dynamics between estimation dates, which is a different question.

The parameter $c$ in \cref{thm:main}(1) does not affect $\A$, since multiplying the functions $k_0(x_j,\cdot)$ by $c$ does not change their linear span. It is a scale parameter of the estimator. Write $k_0=c\,k_0^{(1)}$ with $k_0^{(1)}(x,y)=\e^{b(x+y)}(\e^{axy}-1)$. The RKHS of $k_0$ is that of $k_0^{(1)}$ with $\|h\|_{\HH_0}^2$ divided by $c$, so the estimate \eqref{eq:krr} depends on $c$ and $\lambda$ only through $\lambda/c$. This can also be read off \eqref{eq:beta}, where replacing $k_0$ by $ck_0$ multiplies $K_0$ by $c$ and $\beta$ by $1/c$. Cross-validation of pricing errors therefore determines only this ratio. The role of $c$ becomes visible in the Gaussian process view of KRR, see \cite[Appendix A.6.2]{filipovic3}:
$c$ is the prior variance of the deviation $p-p_0$ from the flat curve $p_0(x)=\e^{bx}$, $\lambda$ is the noise variance, and while the point estimate depends on their ratio only, the posterior variance depends on both.

\subsection{Structure of the hypothesis space}

From now on, $k$ is the kernel of \cref{thm:main}(1). Then $p_0(x)=\e^{bx}$ and, by \eqref{eq:k0},
\begin{equation}\label{eq:k0exp}
  k_0(x,y) = c\,\e^{b(x+y)}\big(\e^{axy}-1\big)=c\,\e^{bx}\big(\e^{(ax+b)y}-\e^{by}\big).
\end{equation}
A direct computation, given in the proof of \cref{thm:main} in \cref{app:proofs}, shows that $k_0(x,\cdot)$ is an eigenfunction of $\alpha$,
\begin{equation}\label{eq:eigen}
  \alpha\big(k_0(x,\cdot)\big)=(ax+b)\,k_0(x,\cdot),\qquad x\ge0 .
\end{equation}
Fix a grid $0<x_1<\dots<x_N$, put $x_0=0$, and let
\begin{equation}\label{eq:lambda}
  \lambda_j = a x_j + b,\qquad j=0,\dots,N .
\end{equation}
Since the $\lambda_j$ are distinct, \eqref{eq:eigen} shows that $k_0(x_1,\cdot),\dots,k_0(x_N,\cdot)$ are linearly independent, so $\dim V=N$. We also write
\begin{equation}\label{eq:phi}
  \phi_j(x) = k_0(x_j,x)/c = \e^{bx_j}\big(\e^{\lambda_jx}-\e^{bx}\big),\qquad j=1,\dots,N,
\end{equation}
for the basis functions of $V$ normalized without the factor $c$.

\begin{lemma}\label{lem:expspan}
The hypothesis space \eqref{eq:A} is the set of functions
\[
  x\mapsto\sum_{j=0}^N z_j\,\e^{\lambda_j x}\qquad\text{with } z\in\R^{N+1},\ \sum_{j=0}^N z_j=1 .
\]
\end{lemma}

The proof is in \cref{app:proofs}. The function $\e^{\lambda_j x}$ is the discount curve of a flat yield curve at level $-\lambda_j$. \cref{lem:expspan} says that the KRR estimate is an affine combination of $N+1$ flat discount curves, whose levels $-b$ and $-(ax_j+b)$ are tied to the grid. For the long end of the curve to remain bounded one needs $\lambda_N=ax_N+b\le0$, that is, $a\le -b/x_N$. When $a$ is sufficiently small relative to the maturity grid, so that $ax_N^2 \ll 1$, we have $\e^{a x_j x}-1\approx ax_jx$ on the grid. The basis functions are then close to collinear, and the kernel matrix $(k_0(x_i,x_j))_{i,j}$ is ill-conditioned. The regularization in \eqref{eq:krr} is essential in this case. 

\begin{proposition}\label{prop:subgrid}
Let $W\subseteq V$ be a linear subspace. Then $p_0+W$ is drift-invariant if and only if
\[
  W=\Span\{k_0(x_j,\cdot): j\in J\}\qquad\text{for some } J\subseteq\{1,\dots,N\}.
\]
\end{proposition}

The proofs of \cref{prop:subgrid} and of \cref{cor:pca} below are in \cref{app:proofs}. \cref{prop:subgrid} shows that the only way to reduce the dimension of $\A$ while keeping drift-invariance is to select a sub-grid of cash flow dates. Such a selection can be made, for example, by a pivoted Cholesky decomposition of the kernel matrix as in \cite{filipovic_kmd}. Reductions based on eigenvectors of the kernel matrix, as in principal component analysis, do not preserve drift-invariance:

\begin{corollary}\label{cor:pca}
Let $K_0=(k_0(x_i,x_j))_{i,j=1}^N$, let $u_1,\dots,u_d\in\R^N$ be eigenvectors of $K_0$ with $1\le d<N$, and let $W$ be the span of the functions $\sum_{j=1}^N u_{l,j}\,k_0(x_j,\cdot)$, $l=1,\dots,d$. Then $p_0+W$ is not drift-invariant.
\end{corollary}

\begin{example}\label{rem:jordan}
\cref{prop:subgrid} describes the drift-invariant affine subspaces of $\A$. There are finite-dimensional drift-invariant affine spaces $p_0+W$ with $W\subseteq C^1(\R_+)$ that are not of this form for any grid. For example, let $a=1$, $b=0$, so that $p_0=1$ and $k_0(x,y)=\e^{xy}-1$, and let $W=\Span\{\e^x-1,\,x\e^x\}$. Then $\alpha(\e^x-1)=\e^x-1$ and $\alpha(x\e^x)=(\e^x-1)+x\e^x$, so $\alpha(W)\subseteq W$, and $p_0+W$ is drift-invariant by \cref{prop:inv}. Here $x\e^x$ is a generalized eigenfunction of $\alpha$. A comparison of Taylor coefficients at \(0\) shows that it cannot be expressed as a linear combination of two functions $\e^{x_1x}-1$ and $\e^{x_2x}-1$. Such spaces correspond to the exponential-polynomial families of \cite{filipovic_exponential}.
\end{example}

\section{Dynamics in the hypothesis space}\label{sec:dyn}

In this section we describe the risk-neutral dynamics of discount curve models with values in the hypothesis space $\A$. Let $(\Omega,\mathcal F,(\mathcal F_t)_{t\ge0},\mathbb Q)$ be a filtered probability space and $W$ a $d$-dimensional Brownian motion. We take $\mathbb Q$ as the risk-neutral measure. Let $k$, the grid, $\lambda_j$ and $\phi_j$ be as in \cref{sec:krr}. We consider discount curve models with values in $\A$, written in the basis \eqref{eq:phi} of $V$,
\begin{equation}\label{eq:model}
  p_t(x) = \e^{bx} + \sum_{j=1}^N Y^j_t\,\phi_j(x),\qquad t,x\ge0,
\end{equation}
where $Y=(Y^1,\dots,Y^N)$ is an It\^o process,
\begin{equation}\label{eq:Y}
  dY_t = \mu_t\,dt + \sigma_t\,dW_t ,
\end{equation}
with progressively measurable $\mu$ and $\sigma$ with values in $\R^N$ and $\R^{N\times d}$, such that the integrals are well defined. We write $\sigma^j_t$ for the $j$-th row of $\sigma_t$. Bond prices are $P(t,T)=p_t(T-t)$ for $t\le T$. The KRR estimate \eqref{eq:beta} corresponds to $Y=c\beta$. The short rate is $r_t=-\partial_xp_t(0)$, which by \eqref{eq:phi} equals
\begin{equation}\label{eq:r}
  r_t = -b - a\sum_{j=1}^N x_j\,\e^{bx_j}\,Y^j_t ,
\end{equation}
and the money-market account is $B_t=\exp\big(\int_0^t r_s\,ds\big)$. The curve $p_t$ lies in $\A$ for all $t$ by construction. We will use the identity
\begin{equation}\label{eq:dk0}
  \phi_j'(x) = \lambda_j\,\phi_j(x) + a x_j\,\e^{b(x_j+x)},
\end{equation}
which follows from \eqref{eq:phi}.

\begin{theorem}\label{thm:dyn}
The following are equivalent:
\begin{enumerate}[(1)]
\item for every $T\ge0$, the discounted bond price $(P(t,T)/B_t)_{t\in[0,T]}$ is a local martingale;
\item for all $j=1,\dots,N$,
\begin{equation}\label{eq:mu}
  \mu^j_t = \big(\lambda_j + r_t\big)Y^j_t = a\Big(x_j-\sum_{i=1}^N x_i\,\e^{bx_i}\,Y^i_t\Big)Y^j_t,\qquad dt\otimes d\mathbb Q\text{-a.e.};
\end{equation}
\item the discounted coefficients $\tilde Y_t=Y_t/B_t$ satisfy
\begin{equation}\label{eq:Ytilde}
  d\tilde Y^j_t = \lambda_j\,\tilde Y^j_t\,dt + \tilde\sigma^j_t\,dW_t,\qquad \tilde\sigma_t=\sigma_t/B_t,\qquad j=1,\dots,N .
\end{equation}
\end{enumerate}
\end{theorem}

\cref{thm:dyn} characterizes the drift of a given It\^o process $Y$. It does not assert that for a given volatility $\sigma$ there exists a process $Y$ with drift \eqref{eq:mu}. This is discussed at the end of this section. The model \eqref{eq:model} is an affine discount term structure model in the sense of \cite[Section 3.2]{filipovic_discount}, and \eqref{eq:mu} is the quadratic drift of \cite[Proposition 3.2]{filipovic_discount} in this case.

\begin{remark}\label{rem:lr}
Let $\tilde Y^0_t=1/B_t$. By \eqref{eq:r}, $d\tilde Y^0_t = -r_t\tilde Y^0_t\,dt = \big(b\,\tilde Y^0_t + a\sum_{j=1}^N x_j\e^{bx_j}\,\tilde Y^j_t\big)dt$. Hence, if the equivalent conditions of \cref{thm:dyn} hold, the process $\tilde Y=(\tilde Y^0,\dots,\tilde Y^N)$ has the linear drift $G\tilde Y$, where $G\in\R^{(N+1)\times(N+1)}$ has entries $G_{00}=b$, $G_{0j}=ax_j\e^{bx_j}$, $G_{jj}=\lambda_j$ for $j\ge1$, and zero otherwise. The discounted bond price is $P(t,T)/B_t=v(T-t)^\top\tilde Y_t$ with $v(x)=(\e^{bx},\phi_1(x),\dots,\phi_N(x))^\top$, and by \eqref{eq:dk0}, $v(x)^\top=e_0^\top\e^{Gx}$, where $e_0=(1,0,\dots,0)^\top$. The equation $d\tilde Y_t=G\tilde Y_t\,dt+\tilde\sigma_t\,dW_t$ has the explicit solution $\tilde Y_t=\e^{Gt}\tilde Y_0+\int_0^t\e^{G(t-s)}\tilde\sigma_s\,dW_s$, where $\tilde\sigma$ is extended by a zero row for the component $\tilde Y^0$. If
\begin{equation}\label{eq:int}
  \E\int_0^T\|\tilde\sigma_t\|^2dt<\infty\qquad\text{for all } T\ge0,
\end{equation}
then $\E[\tilde Y_T\mid\mathcal F_t]=\e^{G(T-t)}\tilde Y_t$ and
\[
  P(t,T) = \frac{e_0^\top\e^{G(T-t)}\tilde Y_t}{\tilde Y^0_t} = \frac{\E[\tilde Y^0_T\mid\mathcal F_t]}{\tilde Y^0_t}>0 .
\]
This is a linear-rational term structure model in the sense of \cite{filipovic_lrtsm}, with state price density $\tilde Y^0$ under $\mathbb Q$. Bond prices are then strictly positive, and discounted bond prices are true martingales. Without \eqref{eq:int}, $p_t$ in \eqref{eq:model} can take negative values.
\end{remark}

\cref{thm:dyn} and \cref{rem:lr} suggest two ways to specify a model in $\A$. The first is to choose the volatility $\sigma$, which may depend on $Y$, and to find a process $Y$ satisfying \eqref{eq:Y} with the drift \eqref{eq:mu}. The drift is quadratic in $Y$, so such a process need not exist for all times, see \cite[Remark 2.8]{filipovic_discount}, and when it exists, $p_t$ need not stay positive. The second is to choose the volatility $\tilde\sigma$ of the discounted coefficients in \eqref{eq:Ytilde}, which may depend on $\tilde Y$. The drift in \eqref{eq:Ytilde} is linear. For an exogenous $\tilde\sigma$ satisfying \eqref{eq:int}, the solution is given explicitly in \cref{rem:lr}. If $\tilde\sigma$ depends on $\tilde Y$, a global solution exists under standard conditions on $\tilde\sigma$, for example a Lipschitz condition. The process $\tilde Y^0$ is then determined by the linear equation $d\tilde Y^0_t=\big(b\,\tilde Y^0_t+a\sum_{j=1}^Nx_j\e^{bx_j}\tilde Y^j_t\big)dt$ with $\tilde Y^0_0=1$. If $\tilde Y^0>0$, then $B=1/\tilde Y^0$ and $Y=\tilde Y/\tilde Y^0$ define a model of the form \eqref{eq:model}--\eqref{eq:Y} satisfying \eqref{eq:mu}, and by \cref{rem:lr} discounted bond prices are true martingales and bond prices are positive. The condition $\tilde Y^0>0$ is the positivity of the state price density, as in \cite{filipovic_lrtsm}. It means that the money-market account $B$ stays finite. It holds in particular if $\tilde Y^j\ge0$ for all $j$. Since $\phi_j\ge0$ by \eqref{eq:phi}, the curve then satisfies $p_t\ge\e^{bx}$, that is, it lies above the flat curve with yield $-b$. For example, geometric Brownian motions $d\tilde Y^j_t=\lambda_j\tilde Y^j_t\,dt+\tilde Y^j_t\,\nu_j^\top dW_t$ with $\nu_j\in\R^d$ satisfy \eqref{eq:int} and $\tilde Y^j>0$, and give a well-defined $N$-factor model.

\section{Conclusion}\label{sec:conclusion}

Kernel ridge regression estimates the discount curve in an affine space $\A$ that is determined by the kernel and by the cash flow dates. We have asked for which kernels this space is invariant under the no-arbitrage drift of the discount curve, for every grid of cash flow dates. Up to kernels with a Hilbert space of dimension at most two, the answer is the family $\e^{b(x+y)}(c\,\e^{axy}+1-c)$ built from the Segal--Bargmann kernel. For this kernel, $\A$ consists of affine combinations of flat discount curves with levels tied to the grid, the drift-invariant subspaces of $\A$ are those of sub-grids, and the discounted coefficients of a risk-neutral model in $\A$ have linear dynamics, which places the model in the linear-rational class and gives well-defined multi-factor models by specifying the volatility of the discounted coefficients.

Two points are left open. The first is empirical: how the kernel \eqref{eq:kernel_intro} performs in curve estimation, and how the sub-grid selection of \cref{prop:subgrid} compares with other dimension reductions. A first step is taken in \cite{celary_rkhs}. The second is the relation between drift-invariance and the smoothness rewards used for static estimation. Whether a weaker form of dynamic consistency, for example for discrete-time dynamics or for a fixed grid, can be combined with a smoothness reward is a question for future work.
%% TO BE DISCUSSED: adjust the open points once the empirical reference is settled.

\subsection*{Acknowledgements}
\noindent\textbf{Funding.} Paul Eisenberg (n\'e Kr\"uhner) acknowledges financial support from the Austrian Science Fund (FWF) under grant PAT4521925.

\medskip
\noindent\textbf{Conflict of interest.}
The authors declare no conflicts of interest.

\medskip
\noindent\textbf{Data availability statement.}
No datasets were generated or analysed during
the current study.

\bibliographystyle{apalike}
\bibliography{bib2}

\clearpage
% Appendices A and B extracted verbatim from main.tex (November 5 source).
% Include with \input{appendices} AFTER \bibliography{bib2} in the main manuscript.
% Keep the \appendix command here; do not duplicate it in main.tex.

\appendix

\section{Interpolating subspaces of an RKHS}\label{app:interp}

This appendix characterizes the set of functions in an RKHS that interpolate given values at finitely many points. The special case of one point is \cref{lem:H0}, which underlies the hypothesis space \eqref{eq:A}. Let $\HH$ be an RKHS with kernel $k$ on a set $\X$. Let $x_1,\dots,x_m\in\X$ be distinct points and $P_1,\dots,P_m\in\R$ target values. Write $\bx=(x_1,\dots,x_m)^\top$, $\bP=(P_1,\dots,P_m)^\top$, and $h(\bx)=(h(x_1),\dots,h(x_m))^\top$ for a function $h:\X\to\R$. We also write $k(\cdot,\bx)=(k(\cdot,x_1),\dots,k(\cdot,x_m))^\top$ and $k(\bx,\bx^\top)=(k(x_i,x_j))_{i,j=1}^m$. We say that $h$ interpolates the target values if $h(\bx)=\bP$. The goal is to characterize the set of interpolating functions in $\HH$,
\[
  \A_{\mathrm{int}} = \{h\in\HH : h(\bx)=\bP\}.
\]
Define the subspaces
\[
  \HH_1 = \Span\{k(\cdot,x_1),\dots,k(\cdot,x_m)\},\qquad \HH_0 = \HH_1^\perp,
\]
so that $\HH=\HH_1\oplus\HH_0$. We denote by $M^+$ the Moore--Penrose pseudoinverse of a matrix $M$.

\begin{lemma}\label{lem:proj}
The orthogonal projections onto $\HH_1$ and $\HH_0$ are given by
\[
  \pi_1 h = k(\cdot,\bx)^\top\, k(\bx,\bx^\top)^+\, h(\bx),\qquad \pi_0 = I_\HH-\pi_1 .
\]
\end{lemma}

\begin{proof}
By the reproducing property, $h(\bx)=\langle k(\cdot,\bx),h\rangle_\HH$. A direct computation shows that $\pi_1$ is idempotent and self-adjoint with range $\HH_1$.
\end{proof}

\begin{lemma}\label{lem:H0gen}
$\HH_0=\{h\in\HH: h(\bx)=\bzero\}$.
\end{lemma}

\begin{proof}
This follows from the reproducing property $h(\bx)=\langle k(\cdot,\bx),h\rangle_\HH$ and the definition of $\HH_0$.
\end{proof}

\begin{lemma}\label{lem:kernels}
$\HH_1$ and $\HH_0$ are RKHS with reproducing kernels
\[
  k_1(x,y) = k(x,\bx^\top)\,k(\bx,\bx^\top)^+\,k(\bx,y),\qquad k_0(x,y)=k(x,y)-k_1(x,y).
\]
\end{lemma}

\begin{proof}
By \cite[Theorem 2.5]{paulsen_rkhs},
a closed subspace $\HH'$ of $\HH$ with orthogonal projection $\pi'$ is an RKHS with kernel $\langle \pi'k(\cdot,x),k(\cdot,y)\rangle_\HH$. For $\HH_1$ this equals $k_1$ by \cref{lem:proj}, and for $\HH_0$ it equals $k-k_1$.
\end{proof}

\begin{lemma}\label{lem:p}
$\A_{\mathrm{int}}\ne\emptyset$ if and only if
\begin{equation}\label{eq:ran}
  \bP\in\ran k(\bx,\bx^\top).
\end{equation}
In this case there is a unique interpolating function $p$ in $\HH_1$. It is given by $p=k(\cdot,\bx)^\top v$, where $v\in\R^m$ is any solution of $k(\bx,\bx^\top)v=\bP$, and $\|p\|_\HH^2=\bP^\top k(\bx,\bx^\top)^+\bP$.
\end{lemma}

\begin{proof}
The first claim is \cite[Theorem 3.4]{paulsen_rkhs}.
Assume \eqref{eq:ran}. A function $q\in\HH_1$ is interpolating if and only if $q=k(\cdot,\bx)^\top w$ for some $w$ with $k(\bx,\bx^\top)w=q(\bx)=\bP$. For two such $v,w$ we have $\|p-q\|_\HH^2=(v-w)^\top k(\bx,\bx^\top)(v-w)=0$. The norm of $p$ is $v^\top k(\bx,\bx^\top)v=\bP^\top k(\bx,\bx^\top)^+\bP$.
\end{proof}

Condition \eqref{eq:ran} holds in particular if $m=1$ and $k(x_1,x_1)>0$.

\begin{lemma}\label{lem:dec}
Assume \eqref{eq:ran} and let $h\in\A_{\mathrm{int}}$. Then $\pi_1h=p$ and
\[
  \|h\|_\HH^2 = \|p\|_\HH^2+\|\pi_0h\|_{\HH_0}^2 .
\]
\end{lemma}

\begin{proof}
We have $(\pi_1h)(\bx)=\langle k(\cdot,\bx),\pi_1h\rangle_\HH=\langle k(\cdot,\bx),h\rangle_\HH=\bP$. Hence $\pi_1h\in\HH_1$ is interpolating, and $\pi_1h=p$ by \cref{lem:p}. The norm identity follows by orthogonality, since $\|\pi_0h\|_\HH=\|\pi_0h\|_{\HH_0}$.
\end{proof}

\begin{theorem}\label{thm:interp}
Assume \eqref{eq:ran}. Then $\A_{\mathrm{int}}=p+\HH_0$, and $\|p+h_0\|_\HH^2=\bP^\top k(\bx,\bx^\top)^+\bP+\|h_0\|_{\HH_0}^2$ for $h_0\in\HH_0$.
\end{theorem}

\begin{proof}
The inclusion $\subseteq$ follows from \cref{lem:dec}. The inclusion $\supseteq$ follows from \cref{lem:H0gen} and the definition of $p$. The norm identity is \cref{lem:dec,lem:p}.
\end{proof}

For $\X=\R_+$, $m=1$, $x_1=0$, $P_1=1$ and $k(0,0)=1$ we get $p=k(\cdot,0)$ and $k_0(x,y)=k(x,y)-k(x,0)k(0,y)$, which is \cref{lem:H0}. For general $m$, \cref{thm:interp} shows that a ridge penalty on the interpolating functions is a penalty on the distance to the minimal norm interpolant $p$, measured in the RKHS $\HH_0$ of the conditional kernel $k_0$.

\section{Proofs}\label{app:proofs}

This appendix collects the proofs of the results of \cref{sec:drift,sec:krr,sec:dyn}, in the order in which they appear.

\subsection{Proof of Proposition~\ref{prop:inv}}
For $h\in V$ a direct computation gives
\begin{equation}\label{eq:eta_decomp}
  \eta(p_0+h) = \eta(p_0) + \alpha(h) - \big(p_0'(0)+h'(0)\big)\,h .
\end{equation}
Since $p_0\in A$ and $p_0+h\in A$, the condition $f(0)=1$ for all $f\in A$ is equivalent to $p_0(0)=1$ and $h(0)=0$ for all $h\in V$.

(1)$\Rightarrow$(2): Taking $h=0$ in \eqref{eq:eta_decomp} gives $\eta(p_0)\in V$. Then $\alpha(h)=\eta(p_0+h)-\eta(p_0)+(p_0'(0)+h'(0))h\in V$ for all $h\in V$.

(2)$\Rightarrow$(1): All three terms on the right-hand side of \eqref{eq:eta_decomp} lie in $V$. This completes the proof of \cref{prop:inv}.

\subsection{Proof of Theorem~\ref{thm:main}}\label{app:proof}
We use that a real analytic function on an interval which vanishes on a set with a limit point in the interval vanishes identically, and that the zeros of a non-constant real analytic function on $\R$ have no limit point.

\begin{lemma}\label{lem:rigid}
Let $k$ be a continuous kernel on $\R_+$. Let $U\subseteq(0,\infty)$ be open and dense in $(0,\infty)$, and let $f,\theta:U\to\R\setminus\{0\}$ be functions such that
\begin{equation}\label{eq:form}
  k(x,y) = f(x)\big(\e^{\theta(x)y}-1\big)\qquad\text{for all } x\in U,\ y\in\R_+ .
\end{equation}
Then either $k=h\otimes h$ for some function $h:\R_+\to\R$, or $k(x,y)=c\,(\e^{axy}-1)$ for some $a,c\in\R\setminus\{0\}$.
\end{lemma}

\begin{proof}
\emph{Step 1.} The set $U\cap\tfrac12U$ is open and dense in $(0,\infty)$, so we can fix $y_1\in U$ with $2y_1\in U$. Put $c_1=f(y_1)$, $\alpha=\theta(y_1)$, $c_2=f(2y_1)$, $\alpha_2=\theta(2y_1)$. By \eqref{eq:form} with $x=y_1$ and $x=2y_1$, and by the symmetry of $k$,
\begin{equation}\label{eq:rows}
  k(x,y_1)=c_1(\e^{\alpha x}-1),\qquad k(x,2y_1)=c_2(\e^{\alpha_2x}-1)\qquad\text{for all } x\in\R_+ .
\end{equation} For $x\in U$ we also have $k(x,2y_1)=k(x,y_1)\big(\e^{\theta(x)y_1}+1\big)$ by \eqref{eq:form}, and $k(x,y_1)\ne0$ since $f(x)\ne0$, $\theta(x)\ne0$. Hence
\begin{equation}\label{eq:R}
  R(x) := \frac{c_2(\e^{\alpha_2x}-1)}{c_1(\e^{\alpha x}-1)}-1,\qquad x\in U,
\end{equation}
satisfies $\e^{\theta(x)y_1}=R(x)$ for all $x\in U$.

The function $R$ extends to a real analytic function on $\R$, with $R(0)=c_2\alpha_2/(c_1\alpha)-1$. We have $R>0$ on $U$, hence $R\ge0$ on $\R_+$. Moreover, $R\ne1$ on $U$, since $k(x,y_1)=f(x)(R(x)-1)\ne0$ for $x\in U$. By \eqref{eq:rows} and \eqref{eq:R},
\begin{equation}\label{eq:ftheta}
  \theta(x)=\frac{\log R(x)}{y_1},\qquad f(x)=\frac{c_1(\e^{\alpha x}-1)}{R(x)-1},\qquad x\in U .
\end{equation}

\emph{Step 2: the case that $R$ is constant.} Then $\theta\equiv\theta_0$ on $U$ for some $\theta_0\ne0$. The symmetry of $k$ on $U\times U$ gives $f(x)(\e^{\theta_0y}-1)=f(y)(\e^{\theta_0x}-1)$, so $f(x)=C(\e^{\theta_0x}-1)$ on $U$ with $C=c_1/(\e^{\theta_0y_1}-1)\ne0$. Hence $k(x,y)=C(\e^{\theta_0x}-1)(\e^{\theta_0y}-1)$ for $x\in U$, $y\in\R_+$, and by continuity and density for all $x,y\in\R_+$. Positive semidefiniteness gives $C>0$, so $k=h\otimes h$ with $h=\sqrt C(\e^{\theta_0\cdot}-1)$.

\emph{Step 3: the case that $R$ is not constant.} We distinguish three cases according to the value $R(0)\in\R_+$.

\emph{Case $R(0)=1$.} By \eqref{eq:ftheta}, $\theta(x)\to0$ as $x\to0$ in $U$, and $0$ is a limit point of $U$. Let $m(x):=f(x)\theta(x)$ for $x\in U$. Put $\ell(z)=(\e^z-1)/z$ for $z\ne0$ and $\ell(0)=1$, which is continuous. By \eqref{eq:ftheta}, $m(x)=c_1(\e^{\alpha x}-1)/(y_1\ell(\theta(x)y_1))$, so
\[
  \lim_{x\to0,\,x\in U}\frac{m(x)}{x} = \frac{c_1\alpha}{y_1} =: \kappa\ne0 .
\]
Fix $y\in U$. By \eqref{eq:form} and symmetry, $x\mapsto k(x,y)=f(y)(\e^{\theta(y)x}-1)$ is differentiable at $0$ with derivative $m(y)$. On the other hand, for $x\in U$, $k(x,y)=m(x)\,y\,\ell(\theta(x)y)$ by \eqref{eq:form}, so $k(x,y)/x\to\kappa y$ as $x\to0$ in $U$. Hence
\begin{equation}\label{eq:mlin}
  m(y)=\kappa\,y\qquad\text{for all } y\in U .
\end{equation}
Now let $x\in U$. By \eqref{eq:form} and \eqref{eq:mlin},
\[  k(x,y_1)=m(x)\,y_1\,\ell(\theta(x)y_1)=\kappa y_1\,x\,\ell(\theta(x)y_1),\]
and by \eqref{eq:rows}, $k(x,y_1)=c_1\alpha\,x\,\ell(\alpha x)=\kappa y_1\,x\,\ell(\alpha x)$. The function $\ell$ is strictly increasing on $\R$, since $\ell'(z)=(\e^z(z-1)+1)/z^2>0$ for $z\ne0$ and $\ell'(0)=1/2$. Hence $\theta(x)y_1=\alpha x$, that is, $\theta(x)=ax$ with $a=\alpha/y_1\ne0$, and $f(x)=m(x)/\theta(x)=\kappa/a=c_1$. Thus $k(x,y)=c_1(\e^{axy}-1)$ for $x\in U$, $y\in\R_+$, and by continuity and density for all $x,y\in\R_+$. This is the second alternative with $c=c_1$.

\emph{Case $R(0)=0$.} Since $R$ is analytic with $R(0)=0$, there are $\varepsilon,C>0$ with $R(x)\le Cx$ for $x\in[0,\varepsilon]$. By \eqref{eq:ftheta}, $f(x)/x\to-c_1\alpha$ as $x\to0$ in $U$. For $y>0$ and $x\in U$ we have $k(x,y)=f(x)(R(x)^{y/y_1}-1)$ by \eqref{eq:form} and \eqref{eq:R}, and $R(x)^{y/y_1}\to0$. Hence $k(x,y)/x\to c_1\alpha$ as $x\to0$ in $U$. For $y\in U$, the function $x\mapsto k(x,y)=f(y)(\e^{\theta(y)x}-1)$ is differentiable at $0$ with derivative $f(y)\theta(y)$. Hence $f(y)\theta(y)=c_1\alpha$ for all $y\in U$, and
\begin{equation}\label{eq:caseB}
  k(x,y)=\frac{c_1\alpha}{\theta(y)}\big(\e^{\theta(y)x}-1\big) = c_1\alpha\Big(x+\frac{\theta(y)}{2}x^2+O(x^3)\Big),\qquad y\in U,\ x\to0 .
\end{equation}
On the other hand, for $x\in U$ and any $y>0$, using $f(x)=c_1\alpha/\theta(x)=c_1\alpha y_1/\log R(x)$,
\[
  k(x,y) = c_1\alpha\,y_1\,\frac{R(x)^{y/y_1}-1}{\log R(x)}
  = c_1\alpha\Big(-\frac{y_1}{\log R(x)} + \frac{y_1R(x)^{y/y_1}}{\log R(x)}\Big).
\]
For $y>2y_1$ the last term is $o(x^2)$ as $x\to0$ in $U$, since $R(x)^{y/y_1}\le C^{y/y_1}x^{y/y_1}$ and $1/\log R(x)\to0$. Comparing with \eqref{eq:caseB} gives, for $y\in U\cap(2y_1,\infty)$,
\[
  -\frac{y_1}{\log R(x)} = x+\frac{\theta(y)}{2}x^2+o(x^2),\qquad x\to0,\ x\in U .
\]
The left-hand side does not depend on $y$. Hence $\theta$ is constant on the non-empty open set $U\cap(2y_1,\infty)$, so $R=\e^{y_1\theta}$ is constant there, and by analyticity on all of $\R$. This contradicts the assumption of Step 3.

\emph{Case $R(0)\notin\{0,1\}$.} Then $R(0)>0$, and we choose $\varepsilon>0$ such that $R(x)>0$ and $R(x)\ne1$ for $|x|<\varepsilon$. Define
\[
  \hat\theta(x)=\frac{\log R(x)}{y_1},\qquad \hat f(x)=\frac{c_1(\e^{\alpha x}-1)}{R(x)-1},\qquad x\in(-\varepsilon,\varepsilon).
\]
Both functions are real analytic, and $\hat\theta=\theta$, $\hat f=f$ on $U\cap(0,\varepsilon)$ by \eqref{eq:ftheta}. We claim that
\begin{equation}\label{eq:hatform}
  k(x,y)=\hat f(x)\big(\e^{\hat\theta(x)y}-1\big)\qquad\text{for all } x\in[0,\varepsilon),\ y\in\R_+ .
\end{equation}
Indeed, for $y\in U$ the function $x\mapsto k(x,y)=f(y)(\e^{\theta(y)x}-1)$ is real analytic on $\R$, the right-hand side of \eqref{eq:hatform} is real analytic on $(-\varepsilon,\varepsilon)$, and both sides agree on the dense subset $U\cap(0,\varepsilon)$ of $(0,\varepsilon)$ by \eqref{eq:form}. Hence they agree on $[0,\varepsilon)$. For general $y\in\R_+$, \eqref{eq:hatform} follows by continuity in $y$ and density of $U$.

By \eqref{eq:hatform} and the symmetry of $k$,
\begin{equation}\label{eq:sym}
  \hat f(x)\big(\e^{\hat\theta(x)y}-1\big)=\hat f(y)\big(\e^{\hat\theta(y)x}-1\big)\qquad\text{for all } x,y\in[0,\varepsilon).
\end{equation}
Write $\beta=\hat\theta(0)=\log R(0)/y_1\ne0$ and $A=\hat f'(0)=c_1\alpha/(R(0)-1)\ne0$, and expand
\[
  \hat f(x)=Ax+f_2x^2+f_3x^3+O(x^4),\qquad \hat\theta(x)=\beta+\theta_1x+\theta_2x^2+O(x^3)
\]
as $x\to0$.
Fix $y\in(0,\varepsilon)$. Comparing the coefficients of $x$, $x^2$ and $x^3$ on both sides of \eqref{eq:sym} gives
\begin{align}
  \hat f(y)\hat\theta(y) &= A(\e^{\beta y}-1), \label{eq:c1}\\
  \tfrac12\hat f(y)\hat\theta(y)^2 &= f_2(\e^{\beta y}-1) + A\theta_1\,y\,\e^{\beta y}, \label{eq:c2}\\
  \tfrac16\hat f(y)\hat\theta(y)^3 &= f_3(\e^{\beta y}-1) + f_2\theta_1\,y\,\e^{\beta y} + A\big(\theta_2y+\tfrac12\theta_1^2y^2\big)\e^{\beta y}. \label{eq:c3}
\end{align}
Since $A\ne0$ and $\beta\ne0$, \eqref{eq:c1} shows $\hat f(y)\hat\theta(y)\ne0$. Let $\psi(y)=y\e^{\beta y}/(\e^{\beta y}-1)$, which extends to a real analytic function on $\R$ with $\psi(0)=1/\beta$. Dividing \eqref{eq:c2} and \eqref{eq:c3} by \eqref{eq:c1} gives
\[
  \hat\theta(y)=\frac{2f_2}{A}+2\theta_1\psi(y),\qquad
  \frac{\hat\theta(y)^2}{6}=\frac{f_3}{A}+\Big(\frac{f_2\theta_1}{A}+\theta_2\Big)\psi(y)+\frac{\theta_1^2}{2}\,y\,\psi(y),
\]
for $y\in(0,\varepsilon)$. Inserting the first identity into the second gives
\begin{equation}\label{eq:quad}
  \frac23\theta_1^2\psi(y)^2-\frac{\theta_1^2}{2}\,y\,\psi(y)+C_1\psi(y)+C_0=0
\end{equation}
with constants $C_0,C_1\in\R$. Both sides of \eqref{eq:quad} are real analytic in $y\in\R$, so \eqref{eq:quad} holds for all $y\in\R$. Let $y\to+\infty$ if $\beta>0$ and $y\to-\infty$ if $\beta<0$. Then $\psi(y)/y\to1$, and dividing \eqref{eq:quad} by $y^2$ gives $(\tfrac23-\tfrac12)\theta_1^2=0$. Hence $\theta_1=0$, and the first identity above shows that $\hat\theta$ is constant on $(0,\varepsilon)$. Then $R=\e^{y_1\hat\theta}$ is constant on $(0,\varepsilon)$, and by analyticity on all of $\R$. This contradicts the assumption of Step 3.
\end{proof}

\begin{lemma}\label{lem:pos}
Let $a,c\in\R\setminus\{0\}$. The function $k(x,y)=c(\e^{axy}-1)$ is a kernel on $\R_+$ if and only if $a>0$ and $c>0$.
\end{lemma}

\begin{proof}
If $a,c>0$ then $k(x,y)=c\sum_{n\ge1}\frac{a^n}{n!}x^ny^n$ is a convergent sum of positive semidefinite kernels, hence positive semidefinite. Conversely, $k(x,x)=c(\e^{ax^2}-1)\ge0$ for all $x$ forces $a$ and $c$ to have the same sign. If $a,c<0$, then the $2\times2$ matrix $(k(x_i,x_j))$ for $x_1=1$ and $x_2=x$ has determinant $c^2\big((1-\e^{a})(1-\e^{ax^2})-(1-\e^{ax})^2\big)$, which converges to $-c^2\e^{a}<0$ as $x\to\infty$. So $k$ is not positive semidefinite.
\end{proof}

We now prove \cref{thm:main}.

\emph{Sufficiency.} In both cases $p_0=k(0,\cdot)=\e^{bx}$ and $\eta(p_0)=0$. In case (1), $k$ is a kernel by \cref{lem:pos} and the fact that products and sums of kernels are kernels. By \eqref{eq:k0},
\[
  k_0(x,y) = c\,\e^{b(x+y)}\big(\e^{axy}-1\big)=c\,\e^{bx}\big(\e^{(ax+b)y}-\e^{by}\big).
\]
For every $x\ge0$ the function $g=k_0(x,\cdot)$ satisfies $g(0)=0$, $g'(0)=c\,a x\,\e^{bx}$ and
\[
  \alpha(g) = g'-g'(0)p_0 = c\,\e^{bx}\big((ax+b)\e^{(ax+b)y} - b\,\e^{by} - ax\,\e^{by}\big) = (ax+b)\,g ,
\]
which is \eqref{eq:eigen}. Hence $\alpha(V)\subseteq V$ for every grid. In case (2), $k$ is a kernel as a sum of two elementary tensors, $k_0(x,y)=\e^{b(x+y)}\varphi(x)\varphi(y)$, and $V\subseteq\Span\{g\}$ with $g(x)=\e^{bx}\varphi(x)$. A direct computation gives $\alpha(g)=(b+\tilde a)g$ if $\varphi(x)=c(\e^{\tilde ax}-1)$ and $\alpha(g)=bg$ if $\varphi(x)=cx$. In both cases every element of $\A$ takes the value $1$ at $0$, so $\A$ is drift-invariant by \cref{prop:inv}.

\emph{Necessity.} Assume that $k$ is drift-invariant. Taking $N=0$, the space $\A=\{p_0\}$ is drift-invariant, so $\eta(p_0)=p_0'-p_0'(0)p_0=0$ and $p_0(0)=1$. Hence $p_0(x)=\e^{bx}$ with $b=p_0'(0)$.

For $x\ge0$ let $g_x=k_0(x,\cdot)$, which is in $C^1(\R_+)$ with $g_x(0)=0$, and $g_0=0$. Taking the grid $\{x\}$ for $x>0$, drift-invariance and \cref{prop:inv} give $\alpha(g_x)\in\Span\{g_x\}$. If $g_x\ne0$ there is $\lambda_x\in\R$ with $g_x'-g_x'(0)\e^{by}=\lambda_xg_x$, $g_x(0)=0$. Here $\delta_x=g_x'(0)\ne0$, since otherwise $g_x$ would solve $g_x'=\lambda_xg_x$, $g_x(0)=0$, and vanish. Solving the linear equation,
\begin{equation}\label{eq:gx}
  g_x(y)=\frac{\delta_x}{b-\lambda_x}\big(\e^{by}-\e^{\lambda_xy}\big)\ \text{ if }\lambda_x\ne b,\qquad g_x(y)=\delta_x\,y\,\e^{by}\ \text{ if }\lambda_x=b .
\end{equation}
In particular, $g_x$ extends to a real analytic function on $\R$ for every $x\ge0$, and by symmetry of $k_0$ so does $k_0(\cdot,y)$ for every $y\ge0$. Define
\begin{align*}
  Z&=\{x\ge0: g_x=0\},\\
  S&=\{x>0: g_x\ne0,\ \lambda_x=b\},\\
  U&=\{x>0: g_x\ne0,\ \lambda_x\ne b\},
\end{align*}
which form a partition of $\R_+$. Note that $k_0$ is a kernel, being the Schur complement of $k$ at $0$, see \cref{lem:kernels}.

\emph{Case 1: $S$ has a limit point in $\R_+$.} For $x,y\in S$, \eqref{eq:gx} and symmetry give $\delta_xy\e^{by}=\delta_yx\e^{bx}$, so $\delta_x\e^{-bx}/x=:c$ does not depend on $x\in S$, and $c\ge0$ since $k_0(x,x)=cx^2\e^{2bx}\ge0$. For fixed $y\in S$, the function $x\mapsto k_0(x,y)-cxy\e^{b(x+y)}$ is real analytic on $\R$ and vanishes on $S$, hence on $\R_+$. Then for fixed $x\ge0$, the function $y\mapsto k_0(x,y)-cxy\e^{b(x+y)}$ is real analytic and vanishes on $S$, hence on $\R_+$. So $k_0(x,y)=cxy\e^{b(x+y)}$ and $k=\e^{b(x+y)}(1+cxy)$, which is case (2) with $\varphi(x)=\sqrt c\,x$.

\emph{Case 2: $Z$ has a limit point in $\R_+$.} For fixed $y\ge0$, the function $x\mapsto k_0(x,y)=g_y(x)$ is real analytic and vanishes on $Z$, hence on $\R_+$. So $k_0=0$ and $k=\e^{b(x+y)}$, which is case (2) with $c=0$.

\emph{Case 3: neither $S$ nor $Z$ has a limit point in $\R_+$.} Then $S\cup Z$ is closed in $\R_+$ and countable, so $U$ is open and dense in $(0,\infty)$. Let $\tilde k(x,y)=\e^{-b(x+y)}k_0(x,y)$, which is a continuous kernel. By \eqref{eq:gx}, for $x\in U$ and $y\in\R_+$,
\[
  \tilde k(x,y)=f(x)\big(\e^{\theta(x)y}-1\big),\qquad f(x)=\frac{\delta_x\e^{-bx}}{\lambda_x-b}\ne0,\qquad \theta(x)=\lambda_x-b\ne0 .
\]
By \cref{lem:rigid}, either $\tilde k=h\otimes h$ for some $h$, or $\tilde k(x,y)=c(\e^{axy}-1)$ with $a,c\ne0$. In the second case, \cref{lem:pos} gives $a,c>0$, and $k=\e^{b(x+y)}(\tilde k+1)$ is of the form (1). In the first case, pick $x\in U$ with $h(x)\ne0$, which exists since $\tilde k(x,x)=h(x)^2\ne0$ for $x\in U$. Then $h(y)=\tilde k(x,y)/h(x)=f(x)(\e^{\theta(x)y}-1)/h(x)$ for all $y\in\R_+$, so $h=\tilde c\,(\e^{\tilde ay}-1)$ with $\tilde a=\theta(x)\ne0$ and some $\tilde c\in\R$. Hence $k=\e^{b(x+y)}\big(1+\tilde c^2(\e^{\tilde ax}-1)(\e^{\tilde ay}-1)\big)$, which is case (2) with $c=|\tilde c|$.

\emph{Dimension.} In case (2), $k=h_1\otimes h_1+h_2\otimes h_2$ with $h_1(x)=\e^{bx}$ and $h_2(x)=\e^{bx}\varphi(x)$, so $k(x,\cdot)\in\Span\{h_1,h_2\}$ for all $x$. Since $\Span\{k(x,\cdot):x\in\R_+\}$ is dense in $\HH$ and $\Span\{h_1,h_2\}$ is finite-dimensional, hence closed, $\HH\subseteq\Span\{h_1,h_2\}$ and $\dim\HH\le2$. In case (1), the functions $k_0(x,\cdot)$, $x>0$, are eigenfunctions of $\alpha$ for the distinct eigenvalues $ax+b$, hence linearly independent, and they lie in $\HH_0\subseteq\HH$ by \cref{lem:H0}. So $\dim\HH=\infty$. The last statement of the theorem follows. This completes the proof of \cref{thm:main}.

\subsection{Proof of Lemma~\ref{lem:expspan}}
By \eqref{eq:k0exp}, $k_0(x_j,\cdot)=c\,\e^{bx_j}\big(\e^{\lambda_j x}-\e^{\lambda_0 x}\big)$ and $p_0=\e^{\lambda_0 x}$. Hence $f\in\A$ if and only if $f=\e^{\lambda_0x}+\sum_{j=1}^N z_j(\e^{\lambda_jx}-\e^{\lambda_0x})$ for some $z_1,\dots,z_N$. The claim follows with $z_0=1-\sum_{j=1}^N z_j$.

\subsection{Proof of Proposition~\ref{prop:subgrid}}
Every $f\in p_0+W$ satisfies $f(0)=1$, and $\eta(p_0)=0\in W$. By \cref{prop:inv}, $p_0+W$ is drift-invariant if and only if $\alpha(W)\subseteq W$. The restriction of $\alpha$ to $V$ is diagonalizable with the eigenbasis $k_0(x_1,\cdot),\dots,k_0(x_N,\cdot)$ and the distinct eigenvalues $\lambda_1,\dots,\lambda_N$, by \eqref{eq:eigen}. The invariant subspaces of a diagonalizable linear map with distinct eigenvalues are exactly the spans of subsets of the eigenbasis.

\subsection{Proof of Corollary~\ref{cor:pca}}
Suppose $p_0+W$ is drift-invariant. By \cref{prop:subgrid} and the linear independence of the $k_0(x_j,\cdot)$, the coefficient space $U_d=\Span\{u_1,\dots,u_d\}$ equals $\Span\{e_j:j\in J\}$ for some $J\subseteq\{1,\dots,N\}$ with $|J|=d$, where $e_1,\dots,e_N$ is the standard basis of $\R^N$. The space $U_d$ is spanned by eigenvectors of $K_0$, so $K_0U_d\subseteq U_d$. For $i\in J$ this gives $K_0e_i\in\Span\{e_j:j\in J\}$, that is, $k_0(x_l,x_i)=0$ for all $l\notin J$. But $k_0(x_l,x_i)=c\,\e^{b(x_l+x_i)}(\e^{ax_lx_i}-1)>0$ for all $l,i\ge1$, and $\{1,\dots,N\}\setminus J$ is not empty. This is a contradiction, and \cref{cor:pca} is proved.

\subsection{Proof of Theorem~\ref{thm:dyn}}
Fix $T$ and write $x=T-t$. It\^o's formula gives
\[
  d\big(P(t,T)/B_t\big)=B_t^{-1}\Big(D_t\,dt + \sum_{j=1}^N \phi_j(x)\,\sigma^j_t\,dW_t\Big),
\]
where
\[
  D_t = -b\,\e^{bx} - \sum_{j=1}^N Y^j_t\,\phi_j'(x) + \sum_{j=1}^N \mu^j_t\,\phi_j(x) - r_t\Big(\e^{bx}+\sum_{j=1}^N Y^j_t\,\phi_j(x)\Big).
\]
Inserting \eqref{eq:dk0} and \eqref{eq:r}, the terms with $\e^{bx}$ cancel and
\[
  D_t = \sum_{j=1}^N \big(\mu^j_t - (\lambda_j+r_t)Y^j_t\big)\,\phi_j(T-t).
\]
(1) holds if and only if $D_t=0$ $dt\otimes d\mathbb Q$-a.e.\ for every $T\ge0$. Since $D_t$ is continuous in $T$, it suffices to take rational $T$, and we obtain one $dt\otimes d\mathbb Q$-null set outside of which $D_t=0$ for all $T\ge t$. This is equivalent to $\sum_j(\mu^j_t-(\lambda_j+r_t)Y^j_t)\phi_j(x)=0$ for all $x\ge0$, $dt\otimes d\mathbb Q$-a.e. By the linear independence of the $\phi_j$, this is (2). The second expression in \eqref{eq:mu} follows from \eqref{eq:lambda} and \eqref{eq:r}.

(2)$\Leftrightarrow$(3): By It\^o's formula, $d\tilde Y^j_t=B_t^{-1}\big(\mu^j_t - r_tY^j_t\big)dt + \tilde\sigma^j_t\,dW_t$, and $\mu^j_t-r_tY^j_t=\lambda_jY^j_t$ is \eqref{eq:mu}. This completes the proof of \cref{thm:dyn}.

\end{document}